%% file: hcpsis.tex
 \documentstyle[prl,aps]{revtex}
\input epsf.tex

\begin{document}
\draft
\twocolumn
\widetext       

\title{Basal-plane incommensurate phases in hcp structures}
\author{I. Luk'yanchuk$^{1,2,*}$, A. Jorio$^1$, M. A. Pimenta$^{1,**}$}
\address{$^1$Departamento de Fisica, Universidade Federal de Minas Gerais, \\
Caixa Postal 702, 30161-970, Belo Horizonte, Minas Gerais, Brazil\\
$^2$L.D.Landau Institute for Theoretical Physics, 117940, Moscow, Russia.}

\date{\today }
\maketitle

\begin{abstract}
 
An Ising model with competing interaction is used to
study the appearance of incommensurate phases  in the basal plane of 
an hexagonal closed-packed structure. 
The calculated mean-field phase diagram reveals
various $1q$-incommensurate and lock-in phases. The results are applied to
explain the basal-plane incommensurate phase in some compounds of the 
$A^{\prime }A^{\prime \prime }BX_4$  family, like $K_2MoO_4$, $K_2WO_4$, $Rb_2WO_4$, 
and to describe the sequence of high-temperature phase transitions in other compounds of this family.
\end{abstract}

\pacs{\leftskip54.8pt PACS: 64.60.Cn, 05.50.+g, 64.70.Rh }


\narrowtext

\setlength{\parindent}{5pt} \leftskip -10pt \rightskip 10pt

\section{INTRODUCTION}

The microscopic origin of incommensurate phases in ferroelectrics, magnetic materials, binary alloys
and other related systems is   subject of interest since the early 60th (for a
review see Ref. \cite{Jan}). It is now well established that incommensurate
modulation in the majority of those systems is caused by the frustrating
competition between  different interatomic or interspin forces responsible for
 structural or magnetic ordering.

The first approach proposed for studying  fru\-stra\-tion-induced
incommensurate phases was based on the axial next-nearest-neighbor Ising (ANNNI) model  with competing uniaxial nearest-neighbor (NN) and
next-nearest-neighbor (NNN) interactions 
in which the structural units  have been described by  binary Ising pseudo-spin
variables \cite{Jan,Selke,Selke2}. Depending
on the coupling parameters and temperature, this model was shown to exhibit a
rich diagram of commensurate and incommensurate modulated phases when the
wave vector jumps between rational and irrational values of the reciprocal lattice period, a phenomenon known as a devil's staircase.

A large variety of incommensurate systems is adequately
described by the ANNNI model  and its analogous \cite{Selke,Selke2,Cum}. The
common feature of these models is that the modulation occurs in the uniaxial
high-symmetry direction. There are, however, examples of incommensurate
phases where wave vector is perpendicular to the high-symmetry axis and can occur in more
than one equivalent direction. The appearance of this kind of incommensurate modulation
was studied for the simple cubic lattice with competing NN and NNN
interactions along the cubic axes and their diagonals \cite{Cub}, for the
simple hexagonal lattice with NN and NNN interactions \cite{Parl1,Parl2,Parl3} and for distorted triangular lattice with only NN interaction \cite{Tri}.

In this paper we  employ the mean-field approximation to study the Ising model
on the hexagonal closed-packed (hcp) lattice, where frustration is
uniquely related to the topology of the lattice and is provided by the 
in-plane NN antiferromagnetic interaction.  We show, for the first time,
that this frustration being stabilized by the  out-plane NN interaction  
 gives rise to   incommensurate phases with the
modulation vector lying in the basal plane. We also study  the phase diagram
of the system when small out-plane NNN interaction is included and show that it
leads to a rich sequence of phase transitions. The used mean-field
consideration  is complementary to the previous cluster-approximation
studies of the hcp Ising model \cite{HI1,HI3} and of the related
hexagonal honeycomb-lattice Ising model \cite{HI1,HI2}. 

Our interest to the hcp Ising model is provided by the transition sequence
in several ionic $A^{\prime }A^{\prime \prime }BX_4$ compounds where the
orientational ordering of $BX_4$ tetrahedra drives a series of structural
phases including  incommensurate states \cite{Cum}. The variety of
transitions can be explained on the basis of the hcp-Ising model where
the orientational states of $BX_4$ tetrahedra are described by two discrete
Ising variables as proposed by Kurzy\'nski and Halawa \cite{Kur,KurRev}. The review of the recent studies of the two-spin  Kurzy\'nski and Halawa model 
is given in \cite{Siems}. We demonstrate that the rigorous treatment of the model explains the
experimentally observed basal-plane incommensurate phases in some $A^{\prime
}A^{\prime \prime }BX_4$ compounds and several other features  not explored in the previous studies.

The paper is organized as follows. In Sect. 2 the structural
properties of $A^{\prime }A^{\prime \prime }BX_4$ family and the motivation
of  the hcp Ising model  are considered. In Sect. 3 we treat
the frustrated hcp Ising model within the mean-field approximation in order to
demonstrate the appearance of  basal-plane incommensurate phases. We  derive
the phase diagram of the system and discuss their relation with the
low-temperature cluster-approximation diagram of \cite{HI1,HI3,HI2}. In Sect. 4 we discuss the application of the obtained
results to the experimental properties of $A^{\prime }A^{\prime \prime }BX_4$
compounds at high temperatures.

Exploring the Ising model, we use the formal terminology of magnetic
systems.  The properties of $A^{\prime }A^{\prime \prime
}BX_4$ compounds are characterized by the electrostatic interaction of $BX_4$
tetrahedra. Consequently, the para-, ferro- and antiferromagnetic terms of
the Ising model  correspond to the para- ferro- and antiferroelectric terms
for $A^{\prime }A^{\prime \prime }BX_4$ compounds. 

\section{COMMENSURATE\ AND\ INCOMMENSURATE\ PHASES IN\ $A^{\prime }A^{\prime
\prime }BX_4$ COMPOUNDS}

The structures and transition sequences in the $A^{\prime
}A^{\prime \prime }BX_4$ family were described in detail in the review
articles \cite{Cum,KurRev}. Shortly, the  $A^{\prime
}A^{\prime \prime }BX_4$ compounds can be presented as a set of $%
BX_4^{2-}$ tetrahedral anions and $A^{\prime \prime +}$ cations that are
regularly placed in the sites of a hcp structure. The $A^{\prime +}$ cations form a simple hexagonal lattice. 

The order-disorder transitions in these compounds  are
related to the  ordering of   degenerate orientations of $BX_4$ tetrahedra in the $A^{\prime}A^{\prime \prime }$ matrix. 
A possible degree of freedom is given by the vertical
up/down orientation of the tetrahedron apexes and an another one by the planar
orientation of the tetrahedra. Orientational ordering of the tetrahedra breaks
the initial hexagonal symmetry $P6_3/mmc$ and leads to a sequence of structural
transitions when temperature decreases. In the Kurzy\'nski and Halawa 
model \cite{Kur}
the vertical and planar orientations are described by  two coupled binary
spin variables, the orientational ordering being provided by the interaction
between neighboring tetrahedra. The ground states of the related two-spins
Ising Hamiltonian  were shown \cite{Kur} to correspond to the experimentally
observed variety of phases in the $A^{\prime }A^{\prime \prime }BX_4$
compounds. There is a common hierarchy in the transition sequence of $A^{\prime
}A^{\prime \prime }BX_4$ compounds. The vertical orientational ordering
occur at higher temperatures (typically of $600-900K$) than the planar one
(below $600K$).

Characteristic feature of $A^{\prime }A^{\prime \prime }BX_4$ family is
the existence of the incommensurate modulations associated with  tetrahedra
orientation that often appear as intermediate phases at structural
transformations. They can be related either to the planar or to the vertical orientational degrees of freedom of the tetrahedra and occur in the low- and in the high-temperature regions, respectively. 

 The incommensurate phases of the first type have been observed 
in a great number of $A^{\prime}A^{\prime \prime }BX_4$ compounds  \cite{Cum,KurRev}. They have the modulation vector directed along the pseudo-hexagonal axis. The appearance of this kind of incommensurate phase was proposed to be related either to specif

ic antisymmetric interaction of the planar orientations of tetrahedra in the unit cell \cite{KurRev,KurCond,Walk,WalkRev} or to 
ANNNI like interaction of the tetrahedra in  neighboring basal planes \cite {Jap}.

The incommensurate phases of the second type are the subject of the present study. The  modulation wave vector associated with the vertical tetrahedra
orientation has the  incommensurate component directed in the basal plane of the hcp structure. This phase occurs in 
alkali molybdates and tungstates $K_2MoO_4$, 
$K_2WO_4$, $Rb_2WO_4$ in the temperature interval of $ 590- 750K$ \cite
{Molyb,MolybX,MolybI}. There is no consistent explanation of this kind of incommensurate phase
although several ideas were proposed in \cite{KurRev}. In this work we explain the appearance of the basal-plane incommensurate structure as a result of
competitive interactions between the vertical orientations of  neighboring
tetrahedra localized in the same basal plane and in the  NN basal planes,
as shown in Fig. 1. 

In the  high-temperature region the planar orientation of the $BX_4$ 
tetrahedra is disordered   and the corresponding planar spin variable  is equal to zero. Thus, in this region, Kurzy\'nski and Halawa model is effectively reduced to the one-spin Ising model where the vertical up/down tetrahedra orientations are described

 by the spin variable $S_i=\pm 1$. The following analysis will be restricted to this region. 

The interaction between tetrahedra has an electrostatic nature. The uniaxial anisotropy of the hcp lattice induces   dipolar moments of $BX_4$ tetrahedra parallel to the $c$-axis.  Therefore, the interaction between the NN tetrahedra is dominanted either 

by their induced dipolar moments or by intrinsic octupolar moments.
  In both   cases the interaction $J_{in}$ between two NN $BX_4$ tetrahedra
localized in the same basal plane favors opposite vertical orientations  
  and therefore has  antiferromagnetic nature. In  contrast, the sign
of the NN out-plane interaction $J_{out}$ depends on  structural details
of the system, like e.g. ratio  $c/a$, effective charges etc.. Thus we consider
both the cases of the ferro- and antiferromagnetic interaction for $J_{out}$.

The mean-field minimization of the free energy given in the next section
shows that the incommensurate structure does exist in a certain region of interaction parameters 
$J_{in}$ and $J_{out}$. It appears, however, that the account of only  NN interaction leads to a degeneracy between   different
phases. To remove this degeneracy, we introduce the weak out-plane
interaction $J_{out}^{\prime }$ between NNN tetrahedra. We show that this
interaction results in   structures that are found experimentally.

\section{MODEL\ AND RESULTS}

We use the Ising spin variable $S_i=\pm 1$ to explore the idea about
frustration induced basal-plane incommensurate phases. The 
Hamiltonian is written as:
\begin{equation}
H=\frac 12\sum_{ij}J_{ij}S_iS_j.  \label{Ising}
\end{equation}
where
\begin{eqnarray}
J_{in} \ \ \ \ \ \ \ \ \text{for the NN {\it in-plane }sites}
 \nonumber \\ 
J_{ij}= \ J_{out} \ \ \ \ \ \ \text{for the NN {\it out-plane }sites} 
\nonumber \\ 
J_{out}^{\prime } \ \ \ \  \text{for the NNN {\it out-plane }sites,}
\label{J}
\end{eqnarray}
as shown on Fig. 1. The negative and positive sign of interaction constants
correspond to ferro- and to antiferromagnetic interactions. As discussed in
the previous Section, we assume that $J_{in}>0$, and that $J_{out}$ and $%
J_{out}^{\prime }$ can take both positive and negative values. It is convenient to introduce 
the driving dimensionless parameters: $\kappa =J_{out}/J_{in}$; $\lambda
=J_{out}^{\prime }/J_{in}$, the last one being assumed to be smaller than
one.

Note  the following properties of the Hamiltonian (\ref{Ising}) that  will be used   later:

\noindent (i) When the ground state for some interaction parameters $\kappa
, $ $\lambda $ is known, the ground state for $-\kappa $, $-\lambda $ is
easily obtained by inversion of signs of the spins in alternated planes.

\noindent (ii) The three-dimensional hcp Ising model can be mapped onto a
two-dimensional honeycomb hexagonal lattice Ising model with NN, NNN, and
NNNN interactions \cite{HI4}.

The ground state of the system is found by the minimization of the free energy:

\begin{equation}
F=-kT\ln Tr\exp (-H/kT). \label{Free}
\end{equation}
 
At $T=0$ the problem reduces
to minimization of the Ising energy $(1/2)\sum J_{ij}S_iS_j$ over all the
possible spin configurations. The effective procedure for solve this problem 
was developed
in \cite{HI1}. Using the mapping honeycomb $\longrightarrow $ hcp lattice
and adopting the results of \cite{HI1} to our variables, we find that six phases: I, II, III, V,
XII and XIII (in notations of \cite{HI1}) whose structures are shown in Fig. 3,
occur at $T=0$. Their energies per one spin are:

\begin{eqnarray}
E_I=J_{in}(3+3\kappa +3\lambda ) \ \ E_{II}=J_{in}(3-3\kappa -3\lambda ) 
\nonumber \\ 
E_{III}=J_{in}(-1+\kappa -3\lambda ) \ \  E_V=J_{in}(-1-\kappa +3\lambda ) 
\nonumber \\ 
E_{XII}=J_{in}(-1+\kappa -\lambda ) \ \ E_{XIII}=J_{in}(-1-\kappa +\lambda ).
\label{Ecom}
\end{eqnarray}
The coexistence lines of these phases (shown in the Fig. 2 by the dots) are defined by 
 the equilibrium of their energies. An important observation given in \cite{HI2} is that the infinite number of other degenerate ground states exists along these lines. These states  become stable at finite temperature due to the entropy factor. The finit

e-temperature cluster-approximation
study shows that the phase sequence can be quite complex (see \cite{HI3} and
Refs. therein).

The low-temperature approach is useful when the values of the spins are
assumed to take a fixed value of either $+1$ or $-1$. Close to $T_c$, the
fluctuations of spins are important and the absolute value of the average $%
\sigma _i=<S_i>$  can be substantially smaller than $1$. To consider
this regime, we minimize the free energy (\ref{Free}) within the
mean-field approximation, the variables $\sigma _i$ being considered as
variational parameters of the model. The standard mean-field treatment \cite
{Bak} gives the following expression for the free energy:

\begin{equation}
F=\frac 12\sum_{ij}J_{ij}\sigma _i\sigma _j+kT\sum_i\int_0^{\sigma _i}\tanh
^{-1}s\;ds.  \label{MF}
\end{equation}

In the vicinity of the transition, the absolute values of $\sigma _i$ are much
smaller than $1$. Expanding (\ref{MF}) in Taylor series we obtain the
Ising-like expression with additional nonlinear terms $\sigma _i^4$:

\begin{equation}
F=\frac{kT}2\sum_i\sigma _i^2+\frac 12\sum_{ij}J_{ij}\sigma _i\sigma _j+%
\frac{kT}{12}\sum_i\sigma _i^4 . \label{F}
\end{equation}
Note that, unlike to the discrete Ising spins $S_i=\pm 1$, the variables $%
\sigma _i$ sweep the continuous spectrum between $+1$ and $-1$. Expression
  (\ref{F}) with arbitrary coefficients is frequently used as a starting
phenomenological functional for considering of   incommensurate phases
in systems with competing interactions \cite{Jan,Parl1,Parl2,Parl3,KurRev}. We use this expression as a basic functional that provides the phase
diagram of the system. The critical temperature of the transition is given by the highest value of $T$ where the functional (\ref{F}) firstly becomes unstable with respect to formation of the  pattern of non-zero $\sigma _i$.  

Let us calculate  the critical temperature and the phase diagram below the
transition as a function of the interaction parameters $\kappa ,\lambda $.
It is convenient to work with the Fourier transformed variables $\sigma
_q=\sum_i\sigma _i\exp (iqr_i)$. Fourier decomposition of (\ref{F}) gives
the following energy per one spin:

\begin{eqnarray}
f=\frac 12\sum_q(kT &+&2J(q))\sigma _q\sigma _{-q}+\frac{kT}4\sum_q(\sigma
_q\sigma _{-q})^2  \nonumber \\
&+&\frac{kT}4\sum_{q\neq q^{\prime }}\sigma _q\sigma _{-q}\sigma _{q^{\prime
}}\sigma _{-q^{\prime }}, \label{Fq}
\end{eqnarray}
where

\begin{eqnarray}
&&\hspace{-0.5cm}J(q)=J_{in}\cdot [\varepsilon ^{\prime }(n_i-n_j)+\kappa
\varepsilon (n_i)\cos q_z+\lambda \varepsilon _c(2n_i)\cos q_z],  \label{Tc}
\\
&&\text{and\thinspace \thinspace \thinspace \thinspace \thinspace \thinspace
\thinspace \thinspace }\varepsilon (n_i)=\cos qn_1+\cos qn_2+\cos qn_3, 
\nonumber \\
&&\hspace{-0.6cm}\varepsilon ^{\prime }(n_i-n_j)=\cos q(n_1-n_2)+\cos
q(n_2-n_3)+\cos q(n_3-n_1) . \nonumber
\end{eqnarray}
We take $n_1=(0,b/3,0)$, $n_2=(a/2,-b/6,0)$, $n_3=(-a/2,-b/6,0)$ as shown in
Fig. 1.

An important assumption was made that expression (\ref{Fq}) contains no
umklapp terms provided by commensurate modulation of the lock-in phases.
Actually, the only phases that give this contribution are these with
modulation vector $q=(0,0,0)$, $(0,0,2\pi /c)$, $(0,2\pi /b,0)$, $(0,2\pi
/b,2\pi /c)$. Their spin configuration $\sigma _j$ corresponds to the states
I, II, III and  V of Fig. 3 with equal in-site amplitudes $\sigma =\left| \sigma
_j\right| $. Free energy of these states will be calculated in a more direct
way later.

The finite-$q$ incommensurate structure becomes stable when the coefficient $%
kT+2J(q)$  is negative. Softening of $kT+2J(q)$ occurs simultaneously in
  several symmetry equivalent points of the $q$-space. The resulting state
is provided by the superposition of the corresponding degenerate plane waves
that can give either a $1q$ stripe phase, or a multi-$q$, double-periodic phase.
Later we will show that the $1q$-phase is more preferable. The modulation
amplitude $\sigma _q$ is a complex value satisfying $\sigma _q=\sigma
_{-q}^{*}$. Since the functional (\ref{Fq}) does not lock its phase we consider 
$\sigma _q$ to be real.

The transition temperature and the modulation vector are defined by
  $kT_{inc}=-\min_{\{q\}}2J(q)$. We found that the $q_c$
component of the modulation vector $q$ is always commensurate with the
reciprocal lattice vector $c^{*}=2\pi /c$ and takes the values of either $0$
or $2\pi /c$. The incommensurate modulation appears in the basal plane along
of either the $a$ or $b$ symmetry directions. Four incommensurate
phases $(q_a,0,0)$, $(q_a,0,2\pi /c)$, $(0,q_b,0)$ and $(0,q_b,2\pi /c)$ are
possible.

The critical temperatures for the phases $(q_a,0,0)$ and $(q_a,0,2\pi /c)$ are
given by

\begin{equation}
kT^\pm_a/J_{in}=\frac{(\kappa -2\lambda )^2}{1\pm 2\lambda }+3 , \label{Tsig}
\end{equation}
where the upper sign corresponds to $q_c=0$ and the lower one to $q_c=2\pi /c$.
The modulation vector

\begin{equation}
q_a=\frac 2a\arccos (\frac{1\pm \kappa }{2\pm 4\lambda })  \label{qa}
\end{equation}
changes from $q_a=0$ (lock-in phases I or II) to $q_a=2\pi /a$ (lock-in
phases V or III).

The critical temperatures for the phases $(0,q_b,0)$ and $(0,q_b,2\pi /c)$ are
given by

\begin{eqnarray}
&k&T^\pm_b/J_{in}=-4(\pm 4\lambda x^4+4x^3\pm (\kappa -2\lambda )x^2
\label{tlam} \\
&&\hspace{1cm}+4(\pm \kappa -3)x\mp (\kappa +\lambda )+1)\qquad  \nonumber \\
&&\hspace{-0.9 cm}\text{where}\hspace{0.2cm}x=\pm ((-3\pm 2\lambda )+[(-3\pm
2\lambda )^2-8\lambda (\kappa \mp 3)]^{1/2})/8\lambda.  \nonumber
\end{eqnarray}
The modulation vector

\begin{equation}
q_b=\frac 2b\arccos x  \label{qb}
\end{equation}
changes from $q_b=0$ (lock-in phases I or II) to $q_b=2\pi /b$ (lock-in
phases V or III).

The incommensurate phases can exist only in that region of parameters $\kappa$ and $\lambda$
when the arguments of $\arccos$ in (\ref{qa}) and in (\ref{qb}) are between $+1$ and $-1$.

Depending on the interaction parameters $\kappa$ and $\lambda$, all four
incommensurate phases can be stable. To find the regions of their stability
one should compare $T_{inc}=T_{a,b}^{\pm }$ with the critical temperatures of
the lock-in phases I, II, III, V, XII and XIII. The later is found directly
from (\ref{F}) since the necessary summation $\sum J_{ij}\sigma _i\sigma _j$
was already performed when calculating (\ref{Ecom}). For the free energy of
the lock-in phases we obtain

\begin{equation}
f_{com}=\frac 12(kT+2E_{com})\sigma ^2+\frac{kT}{12}\sigma ^4,  \label{Fc}
\end{equation}
where $E_{com}$ is the energy (\ref{Ecom}) of corresponding commensurate
phase at $T=0$. The critical temperatures of transitions are given by

\begin{equation}
kT_{com}=-2E_{com}. \label{Tcom}
\end{equation}

After calculation of the maximal critical temperature from $T^\pm_{a,b}$ and $T_{com}$, we  obtain the resulting phase diagram as shown in Fig. 2 by solid lines. The symmetry 
of the diagram with respect to change of sign of the both interaction parameters $\kappa$ and $\lambda$ follows from the property (i) of the Hamiltonian (\ref{Ising}).

To follow the evolution of this diagram when temperature decreases one
should solve the infinite system of the coupled nonlinear variational
equations  obtained from the mean-field functional (\ref{MF}). The
rigorous solution of this problem is beyond the scope of our study.  However,   qualitative aspects obtained in a more simple way are discussed below.

Note, first, that the stripe region of incommensurate phases in Fig. 2 appears at the same 
place where the infinite degenerate lines of phase transitions are predicted
by the cluster calculations at $T=0$ (doted lines). This indicates, 
similarly to the ANNNI model \cite{Selke,Selke2}, the devil's staircase behavior of
the transition sequence below $T_{inc}$. An infinite number of lock-in
phases, where the wave vector jumps between different rational multiples of the reciprocal lattice periods,  appear
at low temperatures. Their phase boundaries converge to the doted lines
at $T=0$. Phases XII and XIII that are stable at low temperatures can appear from  paramagnetic state only via intermediate incommensurate phases. The evolution  of the incommensurate phases when temperature decreases is shown qualitatively in Fig. 4  for

  $J_
{out}^{\prime }=0$. The analogous phase diagram obtained in \cite{Kur}, when the possibility of incommensurate phase was not considered, is shown in the same Figure by the dashed lines.
 
The phase diagram of Fig. 4 is obtained within the approximation when the  nonlinearity of
the functional is modeled by the quartic terms of the Taylor expansion (\ref
{F}) and when the harmonic plane-wave approximation for the modulation
 is used. The incommensurate phase is described by a superposition of $n$ harmonic plane waves with temperature
independent modulation wave vector, corresponding to the symmetry equivalent
points in the $q$-space and with
equal real amplitudes $\sigma _q$. Either stripe phase with $n=1$ or double-modulated phases with $n=2$ or $3$ are possible; in the $3q$-case, $q_1$,$q_2$ and $q_3$  form an   equilateral $120^0$ star. The harmonic
approximation is as exact as $T$ is closer to $T_{inc}$. Variation of $q$ and
contribution of higher harmonics at $T<T_{inc}$  is expected to be small 
 and leads to shift of the phase boundaries of the commensurate low temperature phases towards the lower temerature like it happens in the ANNNI model \cite{Jan,Selke,Selke2}. 

Under these approximations the energy of the incommensurate state (\ref{Fq})
is written as:

\begin{equation}
f_{inc}=nk(T-T_{inc})\sigma _q^2+\frac n2kT\sigma _q^4+n(n-1)kT\sigma _q^4,
\label{Incfun}
\end{equation}
or, after minimization over $\sigma _q$ as:

\begin{equation}
f_{inc=}-\frac 12\frac n{2n-1}\frac kT(T-T_{inc})^2 . \label{Incn}
\end{equation}
Note that (\ref{Incn}) is minimal when $n=1$, i.e., $1q$ is the most stable
incommensurate phase. Comparison of (\ref{Incn}) at $n=1$ with the energy
of corresponding commensurate phase obtained from (\ref{Fc}), by
minimization over $\sigma $

\begin{equation}
f_{com=}-\frac 34\frac kT(T-T_{com})^2 , \label{F2}
\end{equation}
gives the phase sequence below $T_{inc}$. When temperature decreases, the
incommensurate phase is stable up to the temperature

\begin{equation}
T_c=T_{inc}-(3+\sqrt{6})(T_{inc}-T_{com}) , \label{Tfirst}
\end{equation}
defined by the condition $f_{inc}=f_{com}$.   Below $T_c$ a first order transition occurs to   one of the
commensurate phases I, II, III, V, XII, or XIII. The region of 
$\left| \sigma \right| \ll 1$ where the above approximations are valid is
placed above the doted line on Fig. 4. Below this line one can get only
qualitative ideas about the behavior of the transition boundaries. 
 
\section{APPLICATION\ TO $A^{\prime }A^{\prime \prime }BX_4$ COMPOUNDS}

We use now the results of the hcp Ising model with competing interaction to
study the high-temperature phase transitions in the $A^{\prime }A^{\prime
\prime }BX_4$ compounds provided by the vertical orientation of the apexes
of the $BX_4$ tetrahedra. The calculated mean-field phase diagram (Fig. 2)
reveals various incommensurate and lock-in phases that appear just below
the transition and at lower temperatures. Some of the commensurate phases (I, II,
III and V) were  discovered  by Kurzy\'nski and Halawa \cite{Kur}. The occurring
commensurate phases, their correspondence with notation of \cite{Kur}, their
lock-in wave vectors and corresponding symmetry groups are enumerated in  
Table I.

Our calculations reveal the following new features of the phase
diagram. The $1q$ basal-plane incommensurate phases
directed either in the $a$ or $b$ crystallographic direction appear in the
model. The direct first order transition between phases I and
V and between phases II and III is possible when NNN interaction is
included. The new phases XII and XIII can occur at low
temperatures.
   
Table II presents the high-temperature transition sequences for several
typical $A^{\prime }A^{\prime \prime }BX_4$ compounds and their correlation
with the ratio $c/a$. (The data were collected on the basis of \cite{Cum,KurRev,Wyck,PDF}.) Only few compounds reveal the high-symmetry parent 
 phase $P6_3/mmc$ associated with dynamically disordered $BX_4$ tetrahedra. One can imagine
that in the other compounds the phase $P6_3/mmc$ is  virtually present above the
melting point.  

According to their high-temperature transition sequence and to their ratio $c/a$ the $A^{\prime }A^{\prime \prime }BX_4$ compounds can be classified in a following way (see also \cite{KurRev}):

(i) Located in the middle part of the Table II the non underline compounds with
the direct transition (real or virtual) from the phase $P6_3/mmc$ to $Pmcn$ (phase V).
Only a few compounds of this numerous group are given; for other examples 
see: \cite{Cum,KurRev,Wyck,PDF}. Notably, for {\it all} the compounds of this group the ratio $c/a$ varies from $1.27$ to $1.64$.

(ii) Located in the middle part of the Table II the underline compounds where
the phase $P\overline{3}m1$ (phase II) occurs. Two of them,
$Cs_2SO_4$ and $Rb_2SO_4$, demonstrate the reconstructive transition $Pmcn-P%
\overline{3}m1$. Compounds of this group have the similar  ratio $c/a$ as
compounds of (i).

(iii) Located in the upper part of Table II the alkali molybdates and
tungstates $K_2MoO_4$, $K_2WO_4$, $Rb_2WO_4$ \cite{Molyb,MolybX,MolybI} that
 have the intermediate incommensurate phase modulated along
crystallographical direction $b$. The actual transition sequence there is: 
$Pmcn-(0,q_b,2\pi /c)-P6_3/mmc$. To our knowledge, there are no compounds with 
$a$-directed incommensurate phase, although this phase already appears in the NN
approximation. This group has the smallest ratios $c/a\simeq
1.23- 1.25$. The other compounds of this group, $K_2ZnCl_4$, $K_2CoCl_4$, $K_2CoBr_4$, have  similar ratio $c/a$  but the high-temperature phase $P6_3/mmc$ was not reported. 
They can be candidates for the  
basal-plane incommensurate phase if the melting does not
precede the $Pmcn-Inc-P6_3/mmc$ transition.

(iiii) Located in the lower part of Table II compounds that demonstrate the 
 phase $P6_3 $ that is a subgroup of $P6_3mc$ (phase I). Several $A^{\prime }A^{\prime \prime }BeF_4$ compounds and $KLiMoO_4$, with virtual $P6_3-P6_3/mmc$ transition belong to
this group. Two other compounds, $KLiSO_4$ and $RbLiCrO_4$, have the
sequence  $P6_3-Pmcn-P6_3/mmc$ \cite{LiKSO4,KurRev}. These are
particular systems since the complete vertical ordering of the tetrahedra
occurs only in the room-temperature hexagonal phase  $P6_3$. The 
orthorhombic phase $Pmcn$ is characterized by a partial vertical disorder of
tetrahedra. These compounds  reveal the    reconstructive
transition I-V. They can probably be considered as  an intermediate case between classes (i) and (iii). Compounds of the group (iiii) have the largest ratios 
$c/a\simeq 1.64- 1.70$.

Following  Kurzy\'nski and Halawa \cite{Kur}, we assume that 
sign and magnitude of $J_{out}$ (and
therefore of $\kappa $) depend critically on the ratio  $c/a$. 
The  interactions $J_{in}$ and $J_{out}^{\prime }$ are less sensible to variation of $c/a$. Compounds of the groups (i), (iii), (iiii) reproduce the phase diagram of
Fig. 2 if one supposes that $\lambda < -0.5 $ and that $J_{out}$
changes its sign from negative (ferromagnetic) to positive
(antiferromagnetic) when hcp lattice goes from its  expanded along $c$-axis form with $c/a>1.63$ to the contracted form with $c/a<1.63$. It is interesting to observe that the dipolar-dipolar  interaction between two NN out-plane $BX_4$ tetrahedra  changes

 its sign exactly at  $c/a=1.63$. The value of the modulation vector $q_b$ in molybdates and tungstates is correlated with the ratio $c/a$ in the  following way: the smaller the ratio $c/a$ the more $q_b$ deviates from the lock-in value $2\pi /b$ of the p

hase $Pmcn$ \cite{MolybI}. This behavior is consistent with our calculations (\ref{qb}).  Compounds of the group (ii) seem to have a positive $\lambda $ and are placed in the upper right corner of the diagram.

It would be interesting to study the evolution of the phase sequences in 
$A^{\prime }A^{\prime \prime }BX_4$ compounds with a continuous variation of
the interaction parameters. One can achieve this, e.g., by the application of   
uniaxial pressure along the hexagonal axes that slightly changes the ratio $c/a$. In particular, one can expect to obtain 
the commensurate - incommensurate transition, $a$-directed incommensurate
phases and the phase XII with symmetry $Pbca$ (or their subgroups).  
 The classical compound $K_2SeO_4$ from the group (i), having 
the smallest ratio $c/a=1.27$
 could be a good candidate to achieve a tricritical
Lifshitz point and a basal-plane incommensurate phase when submitted to an uniaxial pressure.

Another interesting  result could be  obtained by the application of  an electrical
field $E$ along the hexagonal axis that breaks the mirror-basal-plane
symmetry $S_i\rightarrow -S_i$ in the Hamiltonian (\ref{Ising}). This
results in the additional invariants $E\sum_i\sigma _i$ and $E\sum_i\sigma
_i^3$ in the functional (\ref{F}). The first one slightly favors the
ferroelectric phase I. The second one gives the additional third order term 
$E\sum \sigma _{q1}\sigma _{q2}\sigma _{q3}$ in (\ref{Fq}) where the vectors $%
q_1,q_2,q_3$ form the equilateral triangle (calculations are
analogous to those in \cite{Parl1}). This would lead to the splitting of the
transition $(0,q_b,2\pi /c)-P6_3/mmc$ onto two transitions, with
intervention of the $3q$ modulated incommensurate phase.

In our consideration an interaction  with elastic degrees of freadom that is known to be important in ferroelectrics  has been neglected. 
We expect that this coupling breakes the discontinuity of the transition from  $P6_3/mmc$ to the low temperature phase as it was observed in several $A^{\prime }A^{\prime \prime }BX_4$ compounds. The corresponding analysis is currently in progress. 

We are grateful to N. Speziali for discussion of some experimental
details. The work of I.L. was supported by the Brazilian Agency Fundacao de Amparo a Pesquisa em Minas Gerais (FAPEMIG) and by Russian Foundation of
Fundamental Investigations (RFFI), Grant No. 960218431a.


\input hcpfigs.tex

\end{document}

%% file: hcpfigs.tex
\newpage

\begin{table}
\caption {The lock-in wave vectors $(q_a,q_b,q_c)$  
and space symmetry groups of the commensurate phases appearing in the hcp-Ising model (see also Fig. 3). The correspondence between the notation of Ref.$^a$ used also in this paper and the notations used in Ref.$^b$ is given.}

\begin{tabular}{cccc}
Notations & Notations & Lock-in vector & Symmetry \\
of \tablenotemark[1] & of \tablenotemark[2] & $(q_a,q_b,q_c)$ & group \\ 
\hline
$I$    & $FP$ & $(0,0,0)$             & $P6_3mc$ \\ 
$II$   & $AP$ & $(0,0,2\pi /c)$       & $P\overline{3}m1$ \\ 
$III$  & $CP$ & $(0,2\pi /b,0)$       & $Pmnn$ \\ 
$V$    & $BP$ & $(0,2\pi /b,2\pi /c)$ & $Pmcn$ \\ 
$XII$  & -    & $(\pi /a,0,0)$        & $Pbca$ \\ 
$XIII$ & -    & $(\pi /a,0,2\pi /c)$  & $Pbna$
\end{tabular}
\tablenotetext[1]{ T. Kudo and S. Katsura, Prog. Theor. Phys., {\bf 56}, 435
(1976).}
\tablenotetext[2]{ M. Kurzy\'nski and M. Halawa, Phys. Rev., {\bf B34}, 4846 (1986). }
\end{table}
 

\begin{table}
\caption {The  ratio $c/a$ and the high-temperature sequences 
of transitions  for
several $A^{\prime}A^{\prime \prime }BX_4$ compounds. Question mark 
signifies that either the information about the 
transitions at higher-temperature is not available or the melting  occurs.         The basal-plane incommensurate phase $Inc$ has the modulation vector $(0,q_b,2\pi /c)$.}
\begin{tabular}{ccl}
& $c/a\tablenotemark[1]$ & Transition sequence \\ 
\hline
$K_2ZnCl_4$  & 1.23 & $...-Pmcn-?$ \\ 
$K_2WO_4$    & 1.24 & $...-Pmcn-Inc-P6_3/mmc$ \\  
$K_2CoCl_4$  & 1.24 & $...-Pmcn-?$ \\ 
$K_2CoBr_4$  & 1.24 & $...-Pmcn-?$ \\ 
$K_2MoO_4$   & 1.24 & $...-Pmcn-Inc-P6_3/mmc$ \\ 
$Rb_2WO_4$   & 1.25 & $...-Pmcn-Inc-P6_3/mmc$ \\ 
\hline
$K_2SeO_4$   & 1.27 & $...-Pmcn-P6_3/mmc$ \\ 
$Rb_2MoO_4$  & 1.27 & $...-Pmcn-?$ \\ 
$\underline{ KNaSO_4}$  & \underline{1.29} & 
    $...-\underline{P\overline{3}m1}-?$ \\
$Cs_2WO_4$   & 1.29 & $...-Pmcn-?$ \\ 
$Rb_2SeO_4$  & 1.29 & $...-Pmcn-P6_3/mmc$ \\ 
$K_2SO_4$    & 1.30 & $...-Pmcn-P6_3/mmc$ \\
$Cs_2MoO_4$  & 1.30 & $...-Pmcn-?$ \\ 
$K_2CrO_4$   & 1.30 & $...-Pmcn-?$ \\ 
$K_2MnO_4$   & 1.30 & $...-Pmcn-?$ \\ 
$NaLiBeF_4$  & 1.34 & $...-Pmcn-?$ \\ 
$\underline{Cs_2SO_4}$  & \underline{1.37} 
                 & $...-\underline{Pmcn-P\overline{3}m1}-?$ \\
$\underline{Rb_2SO_4}$ & \underline{1.39} 
                 & $...-\underline{Pmcn-P\overline{3}m1}-?$ \\
$CsLiBeF_4$  & 1.62 & $...-Pmcn-?$ \\ 
$CsLiSO_4$   & 1.62 & $...-Pmcn-?$ \\  
$RbLiSO_4$   & 1.64 & $...-Pmcn-?$ \\
\hline 
$KLiBeF_4$   & 1.64 & $...-P6_3-?$ \\ 
$KLiWO_4$    & 1.65 & $...-P6_3-Cubic$ \\ 
$KLiMoO_4$   & 1.67 & $...-P6_3-Cubic$ \\ 
$TlLiBeF_4$  & 1.68 & $...-P6_3-?$ \\ 
$KLiSO_4$    & 1.69 & $...-P6_3-Pmcn-P6_3/mmc$\\ 
$RbLiBeF_4$  & 1.69 & $...-P6_3-?$ \\ 
$RbLiCrO_4$  & 1.70 & $...-P6_3-Pmcn-P6_3/mmc$
\end{tabular}
\tablenotetext[1]{ Since the ratio $c/a$ for different compounds is given at different temperatures the error bars are estimated as $\pm 0.02$ as the typical variation of thermal expansion. For compounds having an orthorhombic $Pmcn$ symmetry this ratio was estimated as $c/(ab/\sqrt{3})^{1/2}$.}
\end{table}


\newpage
 
\begin{figure}[t]
 
\vspace{0cm}

\hspace{0cm}
\epsfxsize=7truecm
\centerline{\epsffile {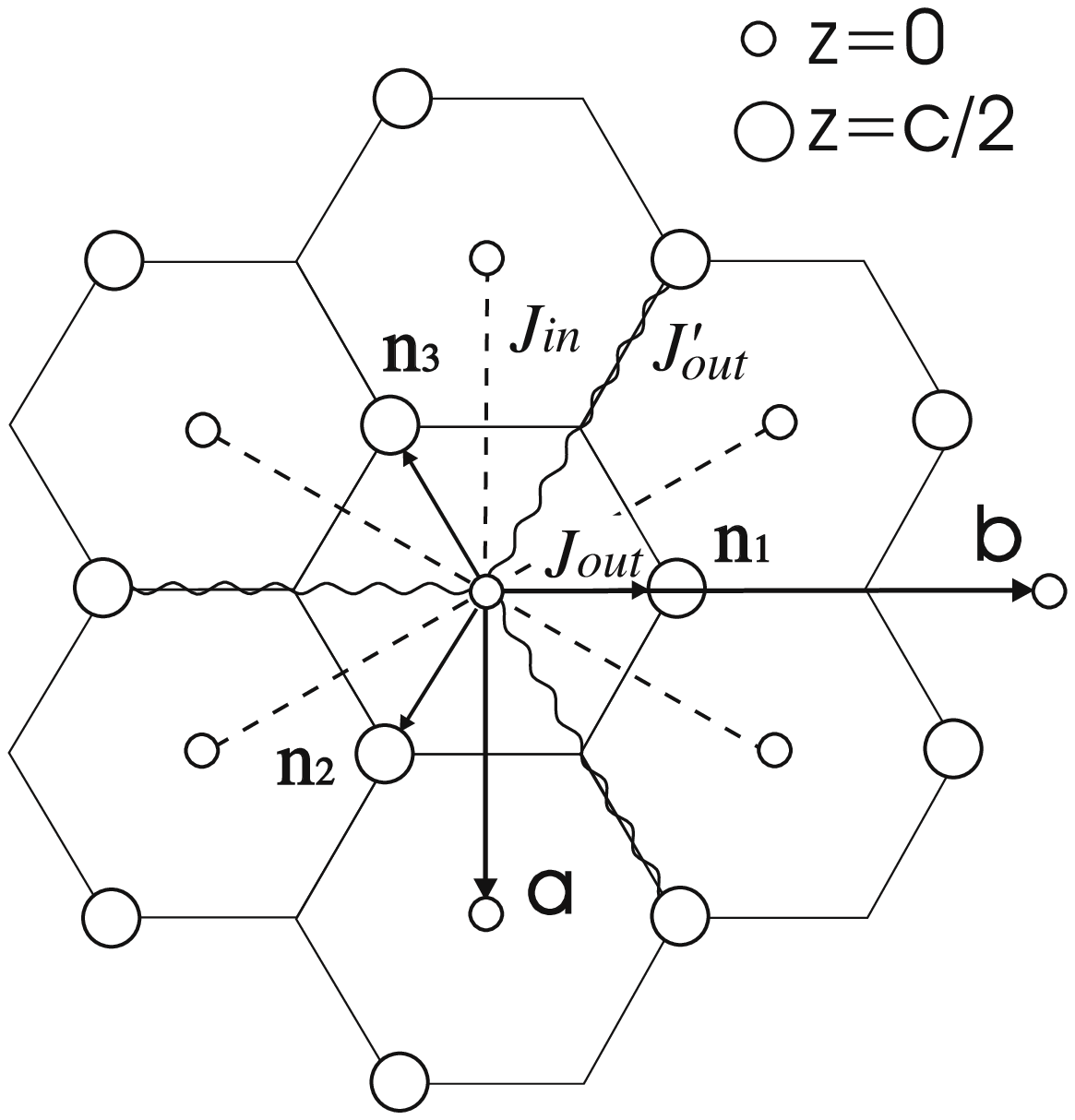}}

\vspace{-4cm}

\caption{$J_{in}$, $J_{out}$ -NN and $J_{out}^{\prime }$ -NNN Ising interactions in hcp structure. We use the orthogonal unit cell ${\bf a},{\bf b,c}$ with 
$b=a\surd{3}$. }

\label{fig1}

\end{figure}


\begin{figure}[t]

\vspace{0cm}

\hspace{0cm}
\epsfxsize=7truecm
\centerline{\epsffile {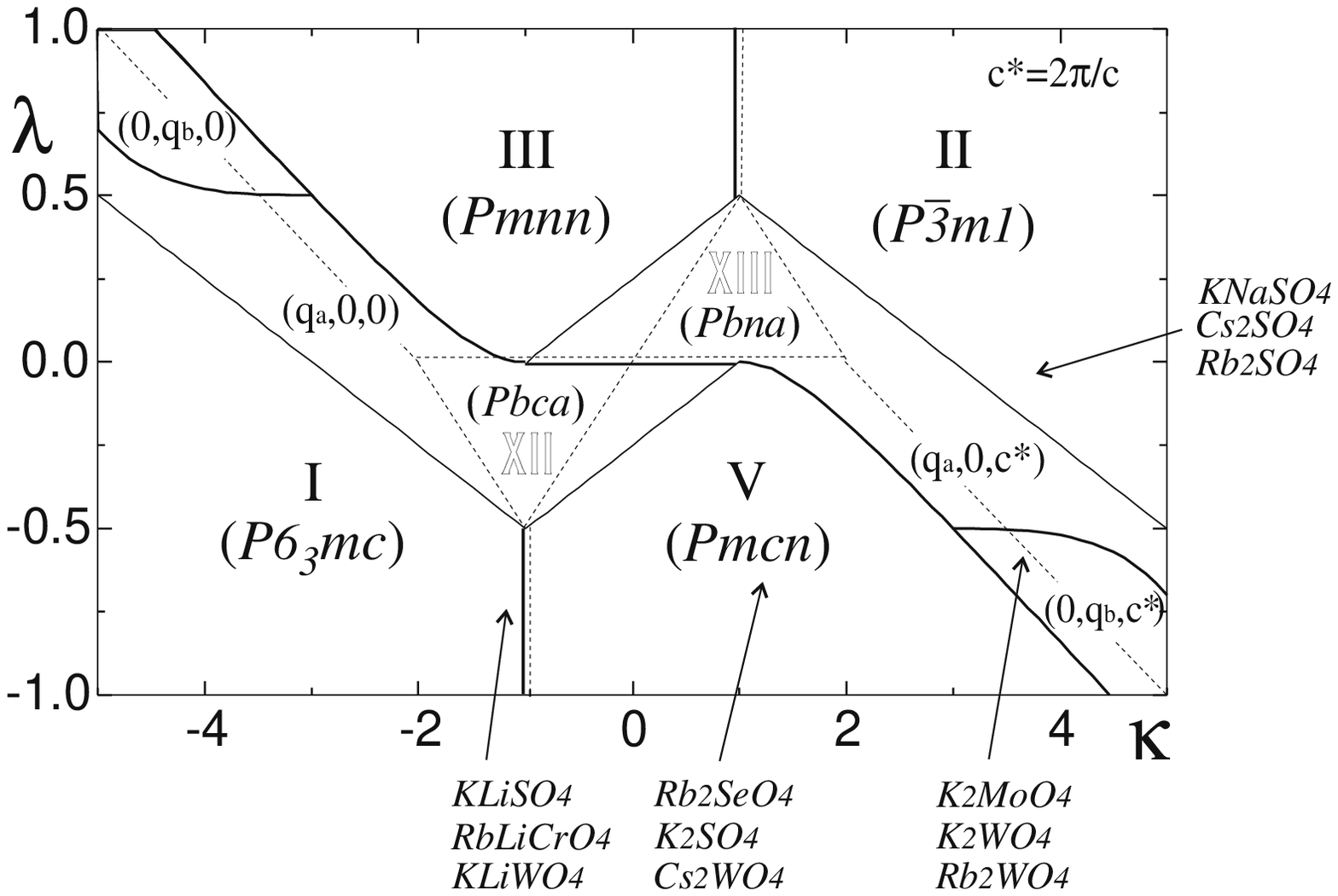}}

\vspace{-4cm}

\caption{ 
Phase diagram of the hcp Ising model as function of the NN and NNN interaction
parameters $\kappa =J_{out}/J_{in}$ and $\lambda =J_{out}^{\prime }/J_{in}$.
Solid lines correspond to the phase diagram just below the transition from the
paramagnetic state. Commensurate phases (roman numbers and corresponding symmetry groups) are also enumerated in Fig. 3 and in Table I. Incommensurate phases are given by their wave vectors $(q_a,q_b,q_c)$.    Doted lines present the phase
diagram at $T=0$. Note that phases  XII and XIII existing at $T=0$ can appear from the paramagnetic states only via an intermediate incommensurate phase. We also show  the possible localization of  some   
$A^{\prime }A^{\prime \prime }BX_4$ compounds. }
 
\label{fig2}
\end{figure}
 
\newpage
 
\begin{figure}[t]
 
\vspace{0cm}

\hspace{0cm}
\epsfxsize=7truecm
\centerline{\epsffile {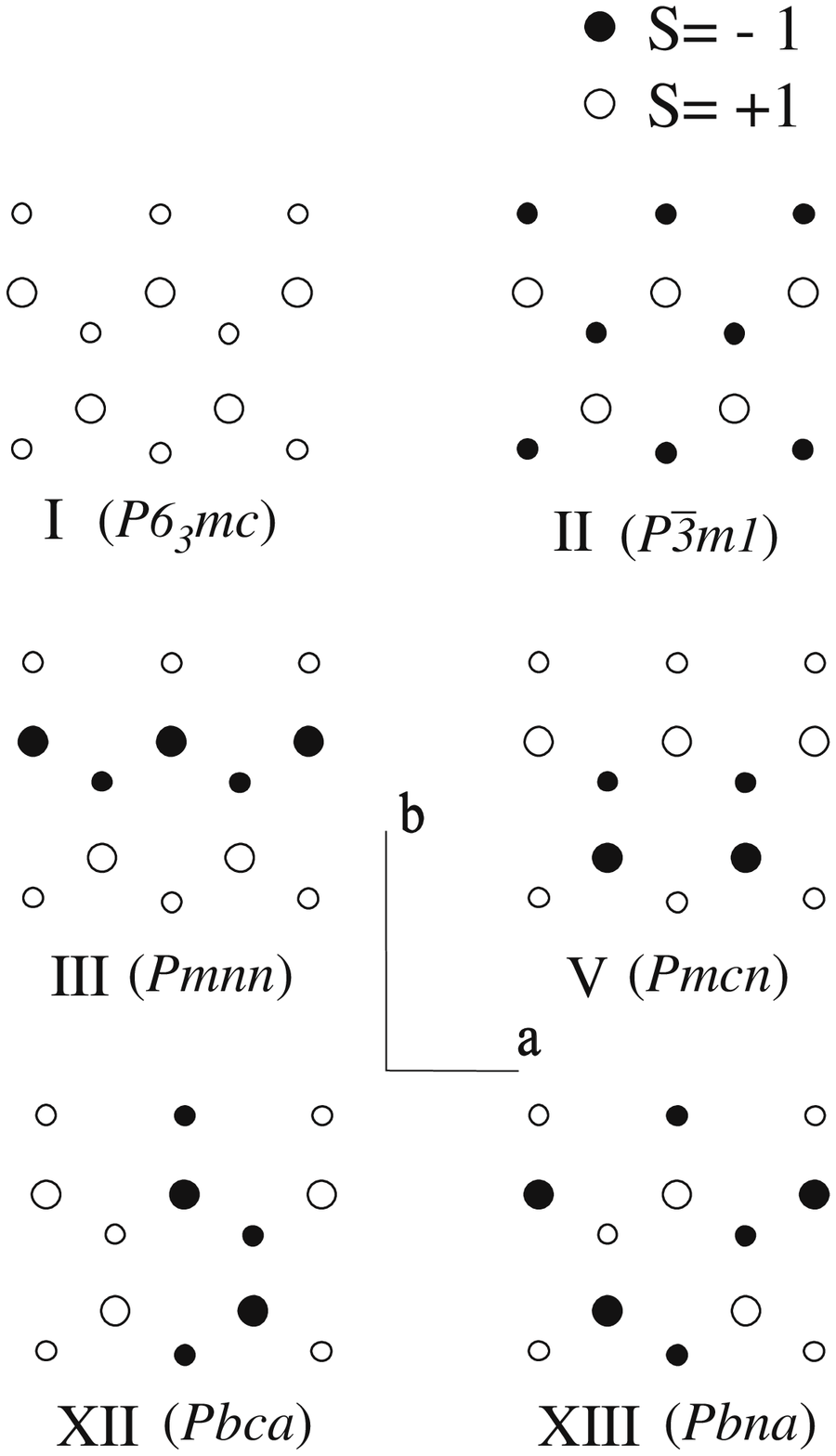}}

\vspace{-0.5cm}

\caption{ 
Spin patterns of the commensurate phases that appear in the hcp Ising model.
Large and small circulus correspond to the spin sites in   alternating
planes of hcp structure. The corresponding lock-in vectors are given in Table I.}
\label{fig3}
\end{figure}

 
\begin{figure}[t]
 
\vspace{0cm}

\hspace{0cm}
\epsfxsize=7truecm
\centerline{\epsffile {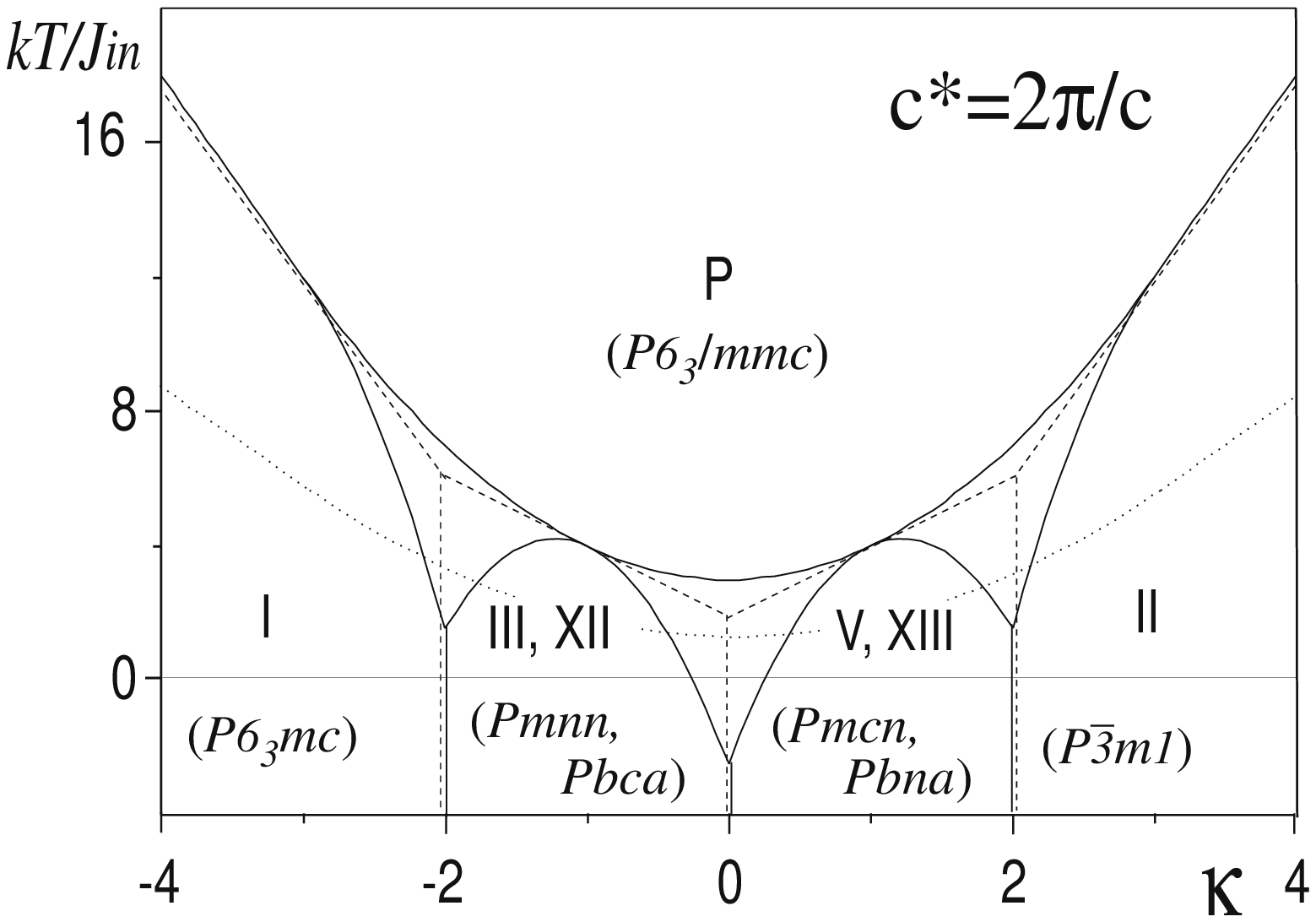}}

\vspace{-4cm}

\caption{Phase diagram of hcp Ising model as function of the NN
interaction parameter $\kappa =J_{out}/J_{in}$ and the reduced temperature $kT/J_{in}$, when the NNN interaction $J_{out}^{\prime }=0$. The transition from the paramagnetic (P) state to the commensurate phase (roman numbers) occurs either directly or via intermediate incommensurate phases $(q_a,0,0)$, $(q_a,0,c^*)$. Dashed lines correspond to the Kurzy\'nski and Halawa (Phys. Rev. {\bf B34}, 4846 (1986)) phase diagram. 
The $\sigma ^4$ expansion of the free energy used for the construction of this
diagram is applicable above the doted line.}

\label{fig4}

\end{figure}
